\documentstyle[12pt]{article}
\begin{document}

\LaTeX{}

\begin{center}
The plasma-solid transition:some implications in
astrophysics\bigskip\bigskip

V.\v Celebonovi\'c $^{1}$ and W.D\"appen$^{2}$

$^{1}$Institute of Physics, Pregrevica 118,11080 Zemun-Beograd, Yugoslavia
\medskip
mail: vladan@phy.bg.ac.yu

$^{2}$Dept. of Physics and Astronomy, USC, Los Angeles,
CA 90089-1342, USA\bigskip
\end{center}
%\begin{center}
%PACS: 64.70 D; 98.38.D
%\end{center}

\begin{center}
key words: plasma-solid transition,interstellar clouds
\end{center}
\bigskip
\begin{center}
to appear in Serb.Astron.J.,    ,XXX,(2002)
\end{center}
\newpage
Abstract: Using a criterion proposed by Salpeter and standard solid-state
physics, we have determined the Debye temperature of a solid in equilibrium
with the electron gas surrounding it. The results obtained can have
astrophysical applications in the determination of parameters of interstellar
and interplanetary clouds.
\bigskip

%Apstrakt: Pomo\'cu kriterijuma koji je predlo\v zio Salpeter,i
%nekih rezultata iz fizike \v cvrstog stanja izra\v cunali smo
%Debajevu temperaturu \v cvrstog tela u ravnote\v zi sa
%elektronskim gasom koji ga okru\v zuje.Rezultati ovog rada mogu da
%nadju primenu u odredjivanju parametara medjuzvezdanih i
%medjuplanetarnih oblaka.

\newpage

\medskip

\begin{center}
Introduction
\end{center}

Phase transitions of various kinds have been, and still are, actively
contributing to important changes occurring in
the world around us. The aim of the present paper
is a contribution to the study of the conditions for the occurrence of
the so-called ``plasma-solid (PS)'' transition.

Simple physical reasoning shows that the PS~transition must
certainly occur in
various astrophysical settings, such as in
proto-planetary and proto-stellar
clouds. On the one hand, it is widely
known that the universe contains different kinds of plasmas.
On the other hand, it is
equally widely known
that solid objects also exist in the universe. This implies that
there must exist a physical region where a transition between
the two regimes takes place.Note also that investigations of the PS transition
could be interesting in non-astrophysical plasma physics, such as in considerations of a laser produced
plasma in front of a metal target ({\it cf.} Dimitrijevi\'c and Konjevi\'c,1980).

In a preliminary study of the PS~transition (\v Celebonovi\'c and D\"appen,
2000), we have determined the conditions for its occurrence using
two simple idealized systems:
a pure Fermi-Dirac and a pure Bose-Einstein gas. The object of the
present calculation is again a determination of the conditions for the
occurrence of the PS~transition, but this time
in a more realistic model. Our starting point is the criterion for the occurrence of the PS~transition
(which was not given that name) proposed by Salpeter (Salpeter,1961).

Salpeter's paper was devoted to a thorough discussion of a zero-tempera-
ture plasma. In the third part of that paper, he considered
a system of positive ions of given
charge and mass, rigidly fixed in the nodes of a perfect crystal
lattice,
and went on to {\it estimate} the zero point energy of the ions.
He showed
that the behavior of such a system can be described by the ratio

\begin{equation}
\label{(1)}\ f_{S}=\frac{E_{z,p}}{E_{C}} \ ,
\end{equation}
%\newpage
where $E_{z,p}$ denotes the zero point energy of the ions and $E_{C}$ is the
Coulomb energy. According to the analysis of (Salpeter,1961) a
PS~transition occurs for $f=1$. Calculations reported in that paper
were estimates, and strictly valid only for $T=0$~K. In the calculations
reported in the following,we have somewhat reformulated this criterion;
namely,we
have compared the energies per particle of a solid and the electron
gas.\newpage
This means that in our formulation,the behaviour of the system is governed by the
ratio

\begin{equation}
 f_{m} = \frac{E_{p,s}}{E_{p,e}}\
\end{equation}
where $E_{p,s}$ denotes the energy per particle of a solid,and $E_{p,e}$
is the energy per particle of the electron gas.Imposing the condition $f_{m}=1$
defines a line (or a region) in the phase space of the system where the
energies per particle in the two phases are equal.In standard phase transition
theory one of the conditions needed for phase equilibrium in a
multi-phase system is the equality of the chemical potentials of its phases (for
example,Landau and Lifchitz,1976). Salpeter's criterion is in fact equivalent to that
statement,but under one supplementary condition - that the energy is
independent on the particle number. Moving off this line (or out of
the region) defined by $f_{m}=1$ implies the occurence of a phase transition.
Our calculations go beyond the original assumptions of Salpeter in at
least two aspects: (i) we have not used estimates but exact
calculations, and (ii) our results take into account the influence of
temperature.
\medskip

%\newpage
\begin{center}
Calculations
\end{center}

Finding a general expression for the energy of a real solid that takes
into account most (if not all) of its characteristics is a formidable
problem in solid state physics (for example Born and Huang ,1968 or Davydov
1980). The complexity of the problem is caused by several factors:the
frequency distribution of the oscillations of particles around their
equilibrium positions,their mutual interactions and their large number (which
is of the order of $10^{23}$).\newpage Various approximations to the complete solution exist;
one of them is the Debye model of a solid. Within the Debye model,
the energy per mole
of a solid is given by the following expression (Born and Huang,1968)

\begin{equation}
\label{(3)}\ E=N\left\{ u\left( v\right) +9nk_{B}T\left[ \left( \frac{T}{\theta }\right)
^{3}\int_{0}^{\theta /T}\left( \frac{1}{2}+\frac{1}{\exp \left( \xi \right)
-1}\right) \xi ^{3}d\xi \right] \right\} \ ,
\end{equation}
%\newpage
where
$N$ denotes the number of elementary cells in a mole of the material,
$n$ the number of particles per elementary cell,
$u(v)=\frac{U}{N}=$ the static lattice energy per cell,
$\theta$ the Debye temperature,
$k_{B}$ the Boltzmann constant, and in the following we will use
the convention $k_{B}=1$.
The second integral in Eq.~(3) can be solved as (Abramowitz and Stegun,1972)

\begin{equation}
\label{(4)}\ I_{1}=\int_{0}^{x}\frac{t^{n}dt}{e^{t}-1}=x^{n}\left[ \frac{1}{n}-\frac{x}{%
2\left( n+1\right) }+\sum_{k=1}^{\infty }\frac{B_{2k}x^{2k}}{\left(
2k+n\right) \left( 2k\right) !}\right]
\end{equation}

under the condition
$ n\succeq 1,\left| x\right| \prec 2\pi $.These conditions are in
(Abramowitz and Stegun, 1972) introduced for purely mathematical
reasons.Physically,
$\left|x\right|\prec 2\pi$ implies $T>\frac{\theta}{2 \pi}$.This
will be used later in this paper,to express the Debye temperature of a solid in
equilibrium with a plasma,as a function of the number density.
Here, the symbol $B_{2k}$ denotes Bernoulli's
numbers. Taking the static lattice energy per cell as the zero point of the
energy scale, it follows from Eq.~(3) that the energy
per particle of a solid within the Debye model is given by
%\newpage
\begin{equation}
\label{(5)}\ E_{p,s}=9T\left( \frac{T}{\theta }\right) ^{3}I
\end{equation}
where
\begin{equation}
\label{(6)}\ I=\frac{1}{2}\int_{0}^{\theta /T}\xi ^{3}d\xi +\int_{0}^{\theta /T}\frac{\xi
^{3}d\xi }{e^{\xi }-1}
\end{equation}
Using Eq.~(4) in Eq.(6), one finally gets

\begin{equation}
\label{(7)}\ \left( \frac{T}{\theta }\right) ^{3}I=\frac{1}{3}+\sum_{k=1}^{\infty }\frac{%
B_{2k}}{\left( 2k+3\right) \left( 2k\right) !}\left( \frac{\theta }{T}%
\right) ^{2k} \
\end{equation}
which implies that

\begin{equation}
\label{(8)}\ E_{p,s}=3T\left[ 1+3\sum_{k=1}^{\infty }\frac{B_{2k}}{(2k+3)\left( 2k\right) !}%
\left( \frac{\theta }{T}\right) ^{2k}\right]
\end{equation}
Expanding explicitly it follows that

\begin{equation}
\label{(9)}\ E_{p,s}\cong 3T\left[ 1+\frac{1}{20}\left( \frac{\theta }{T}\right) ^{2}-\frac{%
1}{1680}\left( \frac{\theta }{T}\right) ^{4}+..\right] \ .
\end{equation}
The energy per particle of a Fermi-Dirac gas of particles of number density
$n$ mass $m$ and spin $s$ is given by (Landau and Lifchitz,1976)

\begin{equation}
\label{(10)}\ E_{p,e}=\frac{gm^{3/2}}{2^{1/2} n \pi ^{2}\hbar ^{3}}F_{3/2}\left( \beta \mu
\right)
\end{equation}
%\newpage
All the symbols in this equation have their usual meaning.In
particular, $g=2s+1$, and $F_{3/2}\left( \beta \mu \right) $
is the $k=3/2$ case of a Fermi integral

\begin{equation}
\label{(11)}\ F_{k}\left( \beta \mu \right) =\int_{0}^{\infty }\frac{\epsilon
^{k}d\epsilon }{1+\exp \left[ \beta \left( \epsilon _{{}}-\mu \right)
\right] } \ .
\end{equation}
Analytical developements of this type of integrals into power series are known
in the literature (such as \v Celebonovi\'c,1998).
The ratio $f_{m}$ as defined in eq.(2) can be formed from eqs.(8) and
(10).After some algebra one gets the following general expression for this
ratio:
\begin{equation}
\label{(12)} f_{m} = (3 2^{1/2} n \pi^{2} \hbar^{3} T)/(gm^{3/2}F_{3/2}\left(
\beta\mu\right))[1+3\sum_{k=1}^{\infty}\frac{B_{2k}}{(2k+3)\left(2k\right)!}%
\left(\frac{\theta}{T}\right)^{2k}]
\end{equation}
This result has the interest of being mathematically general (within a
particular model of a solid).However,due to this generality hardly any
physically useful conclusions can be drawn from it.
Limiting the sum in eq.(12) to terms containing $(\frac{\theta}{T})^{2}$ and inserting
a known power series developement for the Fermi integral leads to the following approximation for the ratio $f_{m}$:
%\newpage
\begin{equation}
\label{(13)}f_{m}=
\frac{9n}{2KT}\frac{20T^2+\theta^2}{(12A^{2}n^{4/3}+5\pi^{2}T^{2})}
\end{equation}
The constant $A$ in the preceeding equation occurs because of the expression
for the chemical potential of the electron gas at $T=0$~K : $\mu _{0}=An^{2/3}$,and it is given by
$A=\left( 3\pi ^{2}\right) ^{2/3}{\hbar ^{2}}/{2m}$.The symbol $K$ denotes the
following combination of constants:
$K=\frac{gm^{3/2}}{2^{1/2}\pi^{2}\hbar^{3}}$.

%\newpage
Imposing the condition $f_{m}=1$ on Eq.(13),
it becomes possible to determine the conditions for the occurence of a PS~transition.
Solving Eq.~(13) for the Debye temperature $\theta $ gives
\begin{eqnarray}
\theta=\frac{1}{3}\surd[2KT/n] \surd[5\pi^{2}T^{2}+12A^{2}n^{4/3}]
\surd[1-\frac{90nT}{(5\pi^{2}T^{2}+12A^{2}n^{4/3})}] \label{(14)}\\ \nonumber
\end{eqnarray}

We have thus obtained the equation of state of a system undergoing a PS
transition. This equation links the relevant parameters of a plasma (number
density and temperature) with those of the solid (Debye
temperature) which is in phase equilibrium with it.

%\newpage
\begin{center}
Discussion
\end{center}
Apart from being interesting from the point of view
of pure statistical physics, this result can find astrophysical
applications in studies of dust and gas clouds. Recent examples of observational
studies of such clouds are for example (Jessop and Ward-Thomson,2000) and
(Gr\"un,Kr\"uger and Landgraf,2000).The particle number density and the temperature of an astrophysical
plasma can be determined from observation using laboratory methods, such as the analysis
of spectral line broadening (e.g.Dimitrijevic,1996) , or in some cases, {\it in-situ}
measurements from space probes.Inserting these values into eq.(14) gives the
possibility to determine the Debye temperature of a solid which is in
equilibrium with the plasma surrounding it.A real physical example of such a
system could be a condensing proto-planetary cloud,such as the one around the
star $\beta$ Pictoris.

Consider Eq.(14) in the limiting case $T\rightarrow0$.Astrophysically,this limit corresponds to an isolated low temperature interstellar cloud,or to
a protoplanetary disk sufficiently distant from the central star.
Mathematically,this means that all functions of $T$ in
eq.(14) can be developed in series with $T$  as a small parameter.Keeping only
the first terms in these developements,one finally gets the following
expression for the Debye temperature:
\begin{equation}
\label{(15)}\ \theta=\frac{2}{3}\surd[2KT/n]3^{1/2}An^{2/3}[1-
\frac{45}{12}\frac{T}{KA^{2}n^{1/3}}(1-
\frac{5}{12}\frac{\pi^{2}T^{2}}{A^{2}n^{4/3}})]
\end{equation}
All the constants occuring in eq.(15) have been previously defined.\newpage
Inserting the values of all the constants which occur in Eq.~(15),and retaining
only the biggest term,one finally arrives at
\begin{equation}
\label{(16)}\ \theta = 67387.5 T^{5/2} n^{-3/2}
\end {equation}
This expression is the final approximation derived in this paper for the
Debye's temperature of a solid in equilibrium with the plasma surrounding
it.Its applicability is limited by the condition
$\frac{\theta}{T}\prec2\pi$,which is built-in in eq.(4).Introducing this
condition in eq.(16) leads to:
\begin{equation}
\label{(17)}T\prec0.00205621\times n
\end{equation}
This is the physical limit of applicability of eq.(16).It is physically
expectable,
because of the limitations inherent to Debye's model on which the
calculation leading to eq.(16) is based.
Expression (16) has two possible applications.
It can be used for the
calculation of $\theta$ if the temperature and the number density of the cloud
are known. Recent observation shows that the temperature in the cold protostellar
clouds in our galaxy is $10<T[{\rm K}]<20$ (Cesarsky,et al.,2000).
%\newpage
Typical values of the
density are in the interval $10^4<n({\rm cm}^{-3}) <10^5$.
Plotting the behaviour of eq.(16) within these limits shows that
$\theta\leq1600 $~K.
In a similar way, the pair of values
$ n = 1.25\times 10^4 {\rm cm}^{-3}$, $T=15$~K would lead to
$ \theta = 630$~K.
%\newpage
Such a
value is only slightly higher than the experimental
value for the element Si.
Note that this last result could be useful for the interpretation
of observations, because emission from silicate grains has indeed
been detected (Cesarsky, et al.,2000)
Turning the argument around,
if solid particles are observed in a cloud,
and if their chemical composition can be determined,
it becomes possible to calculate
the Debye temperature from the principles of solid state physics.
If, in addition, temperature
is known from spectroscopy, one can determine the value of the
chemical potential, and finally, the number density of a cloud from
Eq.~(16).
Further work along these lines is in progress and
will be discussed elsewhere.\\

{\it Acknowledgments}: One of us (W.D.) was
supported in part by the grant AST-9987391 of the
National Science Foundation (USA).
The authors are grateful to Dr.Milan Dimitrijevi\'c of the Astronomical
Observatory in Beograd for helpful comments about an earlier version of the
manuscript.
\newpage
\begin{center}
References
\end{center}

Abramowitz,M. and Stegun,I.A.,1972,Handbook of mathematical
functions, Dover Publications Inc.,New York.

Born,M.and Huang,K.,1968, Dynamical theory of crystal lattices,\newline
OUP,Oxford.

\v Celebonovi\'c,V.,1998, Publ.Astron.Obs.Belgrade,{\bf 60},16.\newline(preprint available at LANL: astro-ph/9802279).

\v Celebonovi\'c,V. and  D\"appen,W.,in:Contributed papers of the 20$^{th}$ SPIG
\newline Conference, (ed. Z.Lj.Petrovi\'c, M.M.Kuraica,
N.Bibi\'c and G.Malovi\'c),p.527,\newline published by the Inst. of Physics,
Faculty of Physics and INN. Vin\v ca,\newline Beograd,Yugoslavia (2000)
(preprint available at LANL: astro-ph/0007337).

Cesarsky,D.,Jones,A.P.,Lequeux,L.,Verstrate,L.,2000,
Astron.Astrophys.,
\newline{\bf 358},708.

Davydov,A.,1980, Th\'eorie du Solide, \'Editions ''Mir'',
Moscou .

Dimitrijevi\'c,M.S. and Konjevi\'c,N, 1980,Optics and Laser Technology.,\newline {\bf 12},145.

Dimitrijevi\'c,M.S.,1996,Zh.Prikl.Spektrosk.,{\bf63},810.

Gr\"un,E.,Krueger,H and Landgraf,M.,2000,in: The Heliosphere at Solar
Minimum:The Ulysses Perspective,eds.Balogh,A.,Marsden,R.,Smith,E.,\newline Springer
Praxis,Heidelberg (in press) (preprint avaliable at LANL:\newline astro-ph/0012226).

Jessop,N.E.and Ward-Thompson,D.,Mon.Not.R.astr.Soc.,\newline (in press)
(preprint avaliable at LANL: astro-ph/0012095)

Landau,L.D.and Lifchitz,E.M., 1976,Statistical Physics,Vol.1,Nauka,\newline
Moscow (in Russian)

Salpeter,E.E.,1961, Astrophys.J., {\bf 134},669.

Ward-Thompson,D.,Andr\'e,P. and Kirk,J.P.,2001,
Mon.Not.R.astr.Soc.\newline(in press)
(preprint avaliable at LANL: astro-ph/0109173)

\end{document}